\documentclass{birkmult}

\newcommand{\Gzero}{G_0}
\newcommand{\dd}{\mathrm{d}}
\newcommand{\ee}{\mathrm{e}}
\newcommand{\bfj}{\mathbf{j}}
\newcommand{\bfJ}{\mathbf{J}}
\newcommand{\bfun}{\mathbf{1}}
\newcommand{\barD}{\bar{D}}
\newcommand{\barG}{\bar{G}}
\newcommand{\barT}{\bar{T}}
\newcommand{\barGzero}{{\bar{G}}_0}
\newcommand{\barSigma}{{\bar{\Sigma}}}
\newcommand{\barZ}{\bar{Z}}
\newcommand{\Hint}{H^\mathrm{int}}

 \newtheorem{thm}{Theorem}[section]

 \theoremstyle{definition}
 
 \theoremstyle{remark}
 \newtheorem{rem}[thm]{Remark}
 
 \numberwithin{equation}{section}

\begin{document}
%
%
%
%
%
%
%
%
%
\title[Structure of Green functions]
 {The structure of Green functions in quantum field theory with
  a general state}
\author[Ch Brouder]{Christian Brouder}

\address{%
Institut de Min\'eralogie et de Physique des Milieux Condens\'es\\
CNRS UMR7590, Universit\'e Paris 6,\\
140 rue de Lourmel, 75015 Paris, France}

\email{christian.brouder@impmc.jussieu.fr}



\date{\today}

\begin{abstract}
In quantum field theory, the Green function is usually calculated
as the expectation value of the time-ordered product of fields
over the vacuum. In some cases, especially in degenerate systems,
expectation values over general states are required. 
The corresponding Green functions are essentially more complex
than in the vacuum, because they cannot be written in terms of
standard Feynman diagrams. Here, a method is proposed to 
determine the structure of these Green functions and to
derive nonperturbative equations for them. The main idea
is to transform the cumulants describing correlations
into interaction terms.
\end{abstract}

\maketitle
\section{Introduction}

High-energy physics uses quantum field theory mainly
to describe scattering experiments through the S-matrix. 
In solid-state or molecular physics, we are rather interested in the 
value of physical observables, such as the charge and current densities
inside the sample or the response to an external perturbation.
At the quantum field theory (QFT) level, these quantities are calculated as
expectation values of Heisenberg operators. For example,
the current density for a system in a state $|\Phi\rangle$
is $\langle \Phi| \mathbf{J}(x) |\Phi\rangle$, where
$|\Phi\rangle$ and $\mathbf{J}(x)$ are written in the
Heisenberg picture.

The first QFT calculation of Heisenberg operators
was made by Dyson in two difficult papers \cite{Dyson51I,Dyson51II}
that were completely ignored.
At about the same time, Gell-Mann and Low discovered that,
when the initial state of the system is nondegenerate, the
expectation value of a Heisenberg operators can be obtained by
a relatively simple formula \cite{GellMann}. The Gell-Mann and Low
formula has been immensely successful and is a key element of the
many-body theory of condensed matter \cite{Fetter,Gross}.
Its main advantage over the formalism developed by
Dyson is that all the standard tools of QFT can be 
used without change.

However, it was soon realized that the assumption of a nondegenerate
initial state is not always valid. As a matter of fact, the problem
of what happens when the initial state is not trivial 
is so natural that it was discussed in many fields of physics:
statistical physics \cite{Fujita}, many-body physics \cite{Hall},
solid-state physics \cite{Esterling},
atomic physics \cite{Lindgren1},
quantum field theory and nuclear physics \cite{Henning,FauserWolter}.
As a consequence, the theory developed to solve this problem 
received several names such as nonequilibrium quantum field
theory (or quantum statistical mechanics) with initial correlations
(or with cumulants, or for open shells, or for degenerate systems). 
It is also called the 
closed-time path or the (Schwinger-)Keldysh approach 
for an arbitrary initial density matrix.

It should be stressed that the problem of the quantum field theory 
of a degenerate system is not only of academic interest. 
For instance, many strongly-correlated systems contain open-shell 
transition metal ions which are degenerate by symmetry. 
This degeneracy makes the system very sensitive to external 
perturbation and, therefore, quite useful for the design of 
functional materials.

The elaboration of a QFT for degenerate systems took a long
time. It started with Symanzik \cite{Symanzik} and Schwinger
\cite{SchwingerJMP} and made slow progress because the
combinatorial complexity is much higher than with standard QFT.
To illustrate this crucial point, it is important to consider an example.
According to Wick's theorem, the time-ordered product
of free fields can be written in terms of normal order
products:
\begin{eqnarray*}
T\varphi(x_1)\dots\varphi(x_4) &=&
{:}\varphi(x_1)\dots\varphi(x_4){:}
+\sum_{ijkl} {:}\varphi(x_i)\varphi(x_j){:}\,\Gzero(x_k,x_l)
\\&&
+\sum_{ijkl} {:}\varphi(x_k)\varphi(x_l){:}\,\Gzero(x_i,x_j)
+\sum_{ijkl} \Gzero(x_i,x_j)\Gzero(x_k,x_l),
\end{eqnarray*}
where the quadruplet of indices $(i,j,k,l)$ runs over
$(1,2,3,4)$, $(1,3,2,4)$ and $(1,4,2,3)$.
The expectation value of this expression over the vacuum
gives the familiar result
$\sum_{ijkl} \Gzero(x_i,x_j)\Gzero(x_k,x_l)$.
However, when the initial state $| \psi\rangle$
is not the vacuum
(as in solid-state physics), we obtain
\begin{eqnarray*}
\langle \psi| T\varphi(x_1)\dots\varphi(x_4) | \psi\rangle
&=&
\langle \psi|{:}\varphi(x_1)\dots\varphi(x_4){:}| \psi\rangle
+\sum_{ijkl} \rho_2(x_i,x_j)\Gzero(x_k,x_l)
\\&&
+\sum_{ijkl} \rho_2(x_k,x_l)\Gzero(x_i,x_j)
+\sum_{ijkl} \Gzero(x_i,x_j)\Gzero(x_k,x_l),
\end{eqnarray*}
where $ \rho_2(x,y)=\langle\psi|{:}\varphi(x)\varphi(y){:}| \psi\rangle$.
If we assume, for notational convenience, that the expectation value of 
the normal product of an odd number of field operators is zero,
the fourth cumulant $\rho_4(x_1,\dots,x_4)$ is defined by the equation
\begin{eqnarray*}
\langle \psi|{:}\varphi(x_1)\dots\varphi(x_4){:}| \psi\rangle
&=&
\rho_4(x_1,\dots,x_4)
+\sum_{ijkl} \rho_2(x_k,x_l)\rho_2(x_i,x_j).
\end{eqnarray*}
If we put $g=\Gzero+\rho_2$, the free four-point Green function
becomes
\begin{eqnarray*}
\langle \psi| T\varphi(x_1)\dots\varphi(x_4) | \psi\rangle
&=&
\rho_4(x_1,\dots,x_4)+ \sum_{ijkl} g(x_i,x_j)g(x_k,x_l).
\end{eqnarray*}
When $\rho_4=0$, the expression is the same 
as over the vacuum,
except for the fact that the free Feynman propagator 
$\Gzero$ is replaced by $g$.
When this substitution is valid, standard QFT can be applied 
without major change and
the structure of the interacting Green functions is not
modified. For fermionic systems described by a 
quadratic Hamiltonian $H_0$, this happens when
the ground state is nondegenerate, so that 
$| \psi\rangle$ is a Slater determinant.
When  $\rho_4\not=0$, the expression becomes essentially
different because the cumulant $\rho_4$ appears as a sort
of free Feynman propagator with four legs. 
In general, the expectation value of a time-ordered product of
$n$ free fields involves $\rho_k$ with $k\le n$.

In other words, the perturbative
expansion of the Green functions can no longer be written 
as a sum of standard Feynman diagrams. 
Generalized Feynman diagrams have to be used,
involving free Feynman propagators with any number
of legs \cite{Fujita,Hall,Kukharenko}.

Because of this additional complexity, the structure of the
Green functions for degenerate systems is almost completely
unknown. The only result available is the equivalent of the
Dyson equation for the one-body Green function 
$G(x,y)$ \cite{Hall}
\begin{eqnarray*}
   G &=& (1-A)^{-1}(\Gzero+C)(1-B)^{-1}(1+\Sigma G),
\end{eqnarray*}
where $A$, $B$, $C$ and $\Sigma$ are sums of one-particle
irreducible diagrams. When the initial state is nondegenerate,
$A=B=C=0$ and the Dyson equation
$G=\Gzero+\Gzero \Sigma G$ is recovered.

In the present paper, a formal method is presented to determine
the structure of Green functions for degenerate systems.
The main idea is to use external sources that
transform the additional propagators $\rho_n$
into \emph{interaction terms}. This brings the
problem back into the standard QFT scheme, where
many structural results are available. 

\section{Expectation value of Heisenberg operators}

Let us consider a physical observable $A(t)$, for instance
the charge density or the local magnetic field.
In the Heisenberg picture, this observable is
represented by the operator $A_H(t)$
and the value of its observable when the system is 
in the state $|\Phi_H\rangle$ is given by the
expectation value
$\langle A (t)\rangle = \langle \Phi_H | A_H(t)|\Phi_H\rangle$.

Going over to the interaction picture, we write the
Hamiltonian of the system as the sum of a free and an
interaction parts: $H(t)=H_0+H_I(t)$, we define the
evolution operator 
$U(t,t')=T\big(\exp(-i\int_{t'}^t H_I(t)\dd t)\big)$
and we assume that the state $|\Phi_H\rangle$
can be obtained as the adiabatic evolution of an
eigenstate $|\Phi_0\rangle$ of $H_0$.
The expectation value of $A$ becomes
\begin{eqnarray*}
\langle A(t) \rangle &=&
  \langle \Phi_0|U(-\infty,t) A(t) U(t,-\infty)|\Phi_0\rangle,
\end{eqnarray*}
where $A(t)$ on the right hand side is the operator representing
the observable in the interaction picture.
The identity $1=U(t,\infty)U(\infty,t)$ and the definition
   $S=U(\infty,-\infty)$ enable us to derive the basic 
expression for the expectation value of an observable
in the interaction picture:
\begin{eqnarray}
\langle A(t)\rangle &=&
  \langle \Phi_0|S^\dagger  T(A(t)S)|\Phi_0\rangle.
\end{eqnarray}
When $|\Phi_0\rangle$ is nondegenerate, this expression can be further
simplified into the Gell-Mann and Low formula
\begin{eqnarray*}
\langle \Phi|A(t)|\Phi \rangle &=&
          \frac{ \langle\Phi_0 | T(A(t)S) | \Phi_0\rangle}
          { \langle\Phi_0 | S | \Phi_0\rangle}.
\end{eqnarray*}

If the system is in a mixed state, as is the case for a
degenerate system by L\"uders' principle, the expectation
value becomes
\begin{eqnarray*}
\langle A(t)\rangle &=&
  \sum_n p_n \langle \Phi_n|S^\dagger  T(A(t)S)|\Phi_n\rangle,
\end{eqnarray*}
where $p_n$ is the probability to find the system in the
eigenstate $|\Phi_n\rangle$.
It will be convenient to use more general mixed states
$\sum_{mn} \omega_{mn} |\Phi_m\rangle \langle \Phi_n|$,
where $\omega_{mn}$ is a density matrix (i.e. a 
nonnegative Hermitian matrix with unit trace).
Such a mixed state corresponds to a linear form
$\omega$ defined by its value over an operator $O$:
\begin{eqnarray*}
\omega(O) &=&
  \sum_{mn} \omega_{mn} \langle \Phi_n|O|\Phi_m\rangle.
\end{eqnarray*}
Then, the expectation value of $A(t)$ becomes
\begin{eqnarray}
\langle A(t)\rangle &=&
  \omega\big(S^\dagger  T(A(t)S)\big).
\label{evAomega}
\end{eqnarray}

\section{QFT with a general state}

In all practical cases, the operator representing the observable
$A(t)$ in the interaction picture is a polynomial in $\varphi$ 
and its derivatives. Its expectation value \eqref{evAomega} 
can be expressed in terms of Green functions that
are conveniently calculated by a formal 
trick due to Symanzik \cite{Symanzik}
and Schwinger \cite{SchwingerJMP}, and reinterpreted by
Keldysh \cite{Keldysh}.

The first step is to define an S-matrix in the presence of an
external current $j$ as
$S(j) = T\big(\ee^{-i\int \Hint(t) \dd t+i\int j(x)\varphi(x)\dd x} \big)$,
where $\Hint$ in the interaction Hamiltonian in the interaction
picture.
The interaction Hamiltonian is then written in terms of a
Hamiltonian density $V(x)$, so that
$\int \Hint(t) \dd t=\int V(x) \dd x$ and
the generating function of the interacting Green functions is
defined by $Z(j_+,j_-)=\omega\big(S^\dagger(j_-)S(j_+)\big)$.
The interacting Green functions can then be obtained
as functional derivatives of $Z$ with respect to the
external currents $j_+$ and $j_-$.
For example
\begin{eqnarray*}
\langle T(\varphi(x)\varphi(y))\rangle =
-  \frac{\delta^2 Z(j_+,j_-)}{\delta j_+(x)\delta j_+(y)},\quad
\mathrm{and}\quad
\langle \varphi(x)\varphi(y)\rangle =
  \frac{\delta^2 Z(j_+,j_-)}{\delta j_-(x)\delta j_+(y)}.
\end{eqnarray*}
As in standard QFT, the connected Green functions are generated
by $\log Z$.

In the functional method \cite{Schwinger,Chou}, the generating
function $Z$ of the interacting system is written as
$Z=\ee^{-iD}Z_0$, where $D$ is the interaction in terms of
functional derivatives 
\begin{eqnarray*}
D &=& \int V\Big(\frac{-i\delta}{\delta j_+(x)}\Big)
  -V\Big(\frac{i\delta}{\delta j_-(x)}\Big) \dd x,
\end{eqnarray*}
and where 
$Z_0(j_+,j_-)=\omega\big(S_0^\dagger(j_-)S_0(j_+)\big)$,
with $S_0(j) = T\big(\ee^{i\int j(x)\varphi(x)\dd x} \big)$.
Note that $Z_0(j_+,j_-)$ is the generating function of the 
free Green functions.

A straightforward calculation \cite{Chou} leads to
\begin{eqnarray*}
Z^0(j_+,j_-) &=& 
  \ee^{-1/2 \int \bfj(x) \Gzero'(x,y) \bfj(y) \dd x \dd y} 
      \ee^{\rho'(j_+-j_-)}, 
\end{eqnarray*}
where $\bfj=(j_+,j_-)$ is the source vector,
\begin{eqnarray}
\Gzero'(x,y) &=& \left( \begin{array}{cc}
    \langle 0 | T\big(\phi(x)\phi(y)\big)|0\rangle
       & -\langle 0 | \phi(y)\phi(x)|0\rangle \\
   -\langle 0 | \phi(x)\phi(y)|0\rangle
        & \langle 0 |
\barT\big(\phi(x)\phi(y)\big)|0\rangle
         \end{array}\right),
\label{defG0}
\end{eqnarray}
is a free Green function (with $\barT$ the anti-time
ordering operator) and
\begin{eqnarray}
\ee^{\rho'(j)} &=& \omega \big( {:}\ee^{i\int j(x)\varphi(x)\dd x}{:}\big)
\label{defrho}
\end{eqnarray}
defines the generating function $\rho'(j)$ of the cumulants
of the initial state $\omega$.

The free Green function $\Gzero'$ describes the dynamics generated
by the free Hamiltonian $H_0$. It can also be written in terms
of advanced and retarded Green functions \cite{SchwingerJMP}.

The idea of describing a state by its cumulants was introduced 
in QFT by Fujita \cite{Fujita} and Hall \cite{Hall}. 
It was recently rediscovered in nuclear
physics \cite{Henning,FauserWolter} and
in quantum chemistry \cite{KutzMukh}.

The next step is to modify the definition of the free
Green function.
The cumulant function is Taylor expanded
\begin{eqnarray*}
\rho'(j) &=& \sum_{n=2}^\infty \frac{1}{n!}
  \int \dd x_1\dots\dd x_n \rho_n(x_1,\dots,x_n) j(x_1)\dots j(x_n).
\end{eqnarray*}
The expansion starts at $n=2$ because $\omega(1)=1$
and the linear term can be removed by shifting the 
field $\varphi$.
The bilinear term $\rho_2(x,y)$ is included into
the free Green function by defining
\begin{eqnarray*}
\Gzero(x,y) &=& \Gzero'(x,y) + \rho_2(x,y) \left( \begin{array}{cc}
    1 & -1 \\ -1 &  1 \end{array}\right),
\end{eqnarray*}
and the corresponding cumulant function becomes
\begin{eqnarray*}
\rho(j) &=& \rho'(j)-(1/2) \int \dd x \dd y j(x) \rho_2(x,y) j(y)\\
 &=& \sum_{n=3}^\infty \frac{1}{n!}
  \int \dd x_1\dots\dd x_n \rho_n(x_1,\dots,x_n) j(x_1)\dots j(x_n).
\end{eqnarray*}

\begin{rem}
There are several good reasons to use 
$\Gzero$ and $\rho$ instead of $\Gzero'$ and $\rho'$:
(i) This modification is exactly what is done in solid-state physics
when the free Green function includes a sum over occupied states
\cite{BrouderPRA};
(ii) At a fundamental level, $\Gzero$ and $\rho$
have a more intrinsic meaning than $\Gzero'$ and $\rho'$
because they do not depend on the state $|0\rangle$ 
chosen as the vacuum; (iii) An important theorem of quantum field
theory \cite{Hollands3}
states that, under quite general conditions, $\rho_n(x_1,\dots,x_n)$
is a smooth function of its arguments when $n>2$, so that 
$\Gzero$ gathers all possible singular terms (a related result was
obtained by Tikhodeev \cite{TikhodeevCor}); (iv) A state
for which $\rho(j)=0$ is called a quasi-free state \cite{Kay1},
quasi-free states are very convenient in practice because
the rules of standard QFT can be used without basic changes.
Thus, the additional complications arise precisely when $\rho$
(and not $\rho'$) is not zero.
\end{rem}

\section{Nonperturbative equations}

To size up the combinatorial complexity due to the presence
of a non-zero $\rho$, we present the diagrammatic expansion
of the one-body Green function $G(x,y)$ for the $\varphi^3$
theory to second order in perturbation theory. For this
illustrative purpose, it will be enough to say that 
the cumulant $\rho_n(x_1,\dots,x_n)$ is pictured as a white 
vertex with $n$ edges attached to it, the other vertex
of the edge is associated with one of the points 
$x_1,\dots,x_n$. For example, $\rho_4(x_1,\dots,x_4)$
is represented by the diagram
\begin{figure}[!ht]
    \includegraphics[width=5.2cm]{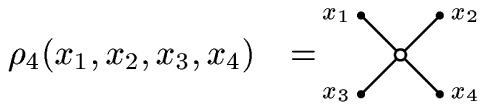}
\end{figure}

In this diagram, the white dot does not stand for a
spacetime point, it just indicates that the points
$x_1$ to $x_4$ are arguments of a common
cumulant.
If we restrict the calculation to the
case when $\rho_n=0$ if $n$ is odd, we obtain the following 
expansion
\begin{figure}[!ht]
    \includegraphics[width=12cm]{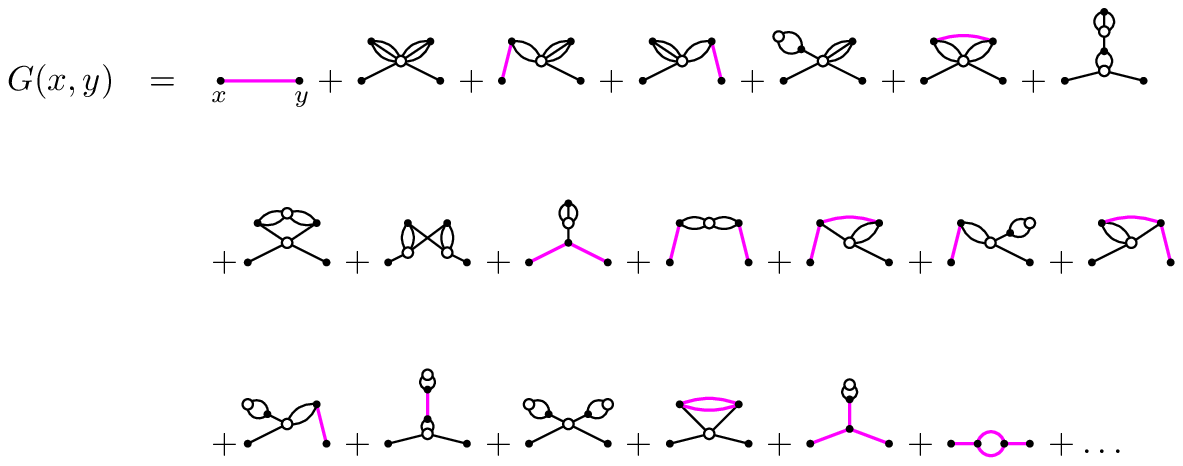}
\end{figure}

In standard QFT, only the first and last diagrams of the right hand
side are present. In the general case when all $\rho_n\not=0$,
the number of diagrams is still much larger. 

\subsection{Generalized Dyson equation}
As mentionned in the introduction, the only known result concerning
the structure of Green functions with a general state was derived
by Hall for the one-body Green function $G(x,y)$ \cite{Hall}
\begin{eqnarray*}
   G &=& (1-A)^{-1}(\Gzero+C)(1-B)^{-1}(1+\Sigma G).
\end{eqnarray*}
In diagrammatic terms the quantities $A$, $B$,
$C$ and $\Sigma$ are sums of one-particle irreducible
diagrams. If we take our example of the Green function
of $\varphi^3$ theory up to second order, we find
\begin{figure}[!ht]
   \includegraphics*[width=7cm]{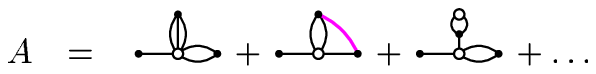}
\end{figure}
\begin{figure}[!ht]
   \includegraphics*[width=7cm]{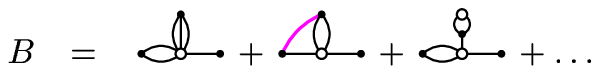}
\end{figure}
\begin{figure}[!ht]
   \includegraphics*[width=12cm]{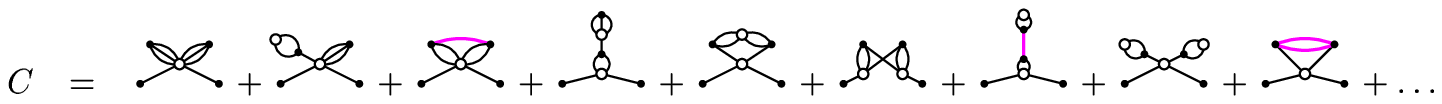}
\end{figure}
\begin{figure}[!ht]
   \includegraphics*[width=7cm]{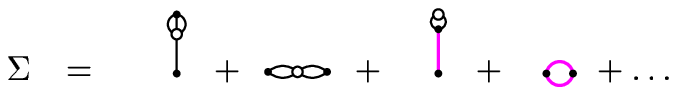}
\end{figure}

In standard QFT, we have $A=B=C=0$ and the diagrammatic
representation of $\Sigma$ contains much less terms.
However, the difference with standard QFT is not only limited to 
the number of diagrams. The definition \eqref{defrho} of the
cumulant function, and the fact that the free field $\varphi$
is a solution of the Klein-Gordon equation imply that
$\rho_n$ is a solution of the Klein-Gordon equation
in each of its variables. Thus, $A(x,y)$, $B(x,y)$
and $C(x,y)$ are solutions of the Klein-Gordon equation
for $x$ and $y$. As a consequence,
applying the Klein-Gordon operator to the Green function
gives us $(\Box+m^2)G=(1-B)^{-1}(1+\Sigma G)$.
In other words, applying the Klein-Gordon operator kills
a large number of terms of $G$. This is in stark contrast
with standard QFT, where $(\Box+m^2)G=1+\Sigma G$ and
amputating a Green function does not modify its structure.
This important difference makes some tools of standard QFT
(e.g. amputated diagrams or Legendre transformation)
invalid in the presence of a general state.

All those difficulties explain the scarcity of results 
available in non-perturbative QFT with a general state.
Apart from Hall's work \cite{Hall}, the only non-perturbative
results are Tikhodeev's cancellation theorems
\cite{Tikhodeev,Danielewicz}
and the equation of motion for the Green 
functions \cite{BrouderEuroLett}.

In the next section, we present a simple trick to derive
the structure of Green functions with a general state.

\subsection{Quadrupling the sources}
We first determine the main formal
difference between standard QFT and QFT with a general state.
In both cases, the generating function of the Green functions can be
written $Z=\ee^{-iD}Z_0$, where $D$ describes the interaction
and $Z_0$ the initial state. In the presence of a general state,
the interaction $D$ is simple but $Z_0$ is
made non standard by the cumulant factor $\ee^{\rho}$.
The idea of the solution is to
transfer the cumulant function $\rho$ from $Z_0$ to $D$,
because powerful functional methods were
developed to deal with general interactions $D$. These methods
were first proposed by Dominicis and Englert \cite{DominicisEnglert}
and greatly expanded by the 
Soviet school \cite{Vasilev1,Vasilev2,Vasilev3,Vasilev4,%
Vasilev,Pismak1,Pismak2,Pismak3,Pismak4}. 

This transfer from the initial state to the interaction
can be done easily by introducing two additional external
sources $k_+$ and $k_-$ and using the identity
 \begin{eqnarray*}
  \ee^{\rho(j_+-j_-)} = 
  \ee^{\rho(-i\frac{\delta}{\delta k_+}-i\frac{\delta}{\delta k_-})}
  \ee^{i\int (j_+(x)k_+(x)-j_-(x)k_-(x))\dd x}\big|_{k_+=k_-=0}.
 \end{eqnarray*}
The term involving $\rho$ can now be transferred from
$Z_0$ to $D$ by defining the new generating function
\begin{eqnarray*}
\barZ(j_\pm,k_\pm) &=& \ee^{-i\barD} \barZ_0(j_\pm,k_\pm),
\end{eqnarray*}
where the modified interaction is
\begin{eqnarray*}
\barD &=& \int V\Big(\frac{-i\delta}{\delta j_+(x)}\Big)
  -V\Big(\frac{i\delta}{\delta j_-(x)}\Big) \dd x
  -i\rho(-i\frac{\delta}{\delta k_+}-i\frac{\delta}{\delta k_-}),
\end{eqnarray*}
and the modified free generating function is
\begin{eqnarray*}
\barZ_0(j_\pm,k_\pm) &=&
           \ee^{-1/2 \int \bfJ(x) \barGzero(x,y) \bfJ(y)\dd x \dd y},
\end{eqnarray*}
with $\bfJ=(j_+,j_-,k_+,k_-)$.
The modified free Green function $\barGzero$ is now a 4x4 matrix
that can be written as a 2x2 matrix of 2x2 matrices
\begin{eqnarray*}
\barGzero &=& \left( \begin{array}{cc}
    \Gzero
       & -i\bfun \\
   -i\bfun
        & 0 \end{array}\right).
\end{eqnarray*}
In contrast to the standard case, the free Green function
$\barGzero$ is invertible
\begin{eqnarray*}
{\barGzero}^{-1} &=& \left( \begin{array}{cc}
    0
       & i\bfun \\
   i\bfun
        & \Gzero \end{array}\right),
\end{eqnarray*}
and it is again possible to use amputated diagrams and Legendre
transformations. The free generating function $\barZ_0$ is 
the exponential of a function that is bilinear in the sources,
and all the standard structural tools of QFT are available again.
We illustrate this by recovering Hall's analogue of the Dyson equation.

\subsection{An algebraic proof of Hall's equation}
The free generating function $\barZ_0$ has a standard form and
the Dyson equation holds again:
$\barG=\barGzero+\barGzero\barSigma\barG$, where
$\barG$ is the 4x4 one-body Green function obtained from
the generating function $\barZ$ and $\barSigma$ is
the corresponding self-energy.
Each 4x4 matrix is written as a 2x2 matrix of 2x2 matrices.
For example
\begin{eqnarray*}
\barG &=& \left( \begin{array}{cc}
    \barG_{11}
       & \barG_{12} \\
   \barG_{21}
        & \barG_{22} \end{array}\right).
\end{eqnarray*}
We want to determine the structure of the 
2x2 Green function $G$, which is equal to
$\barG_{11}$ when $k_+=k_-=0$.

The upper-left component of the Dyson equation for $\barG$ is
\begin{eqnarray}
\barG_{11} &=& \Gzero + (\Gzero\barSigma_{11}-i\barSigma_{21})
  \barG_{11}+ (\Gzero\barSigma_{12}-i\barSigma_{22})\barG_{21}.
\label{upperleft}
\end{eqnarray}
The lower-left component gives us
$\barG_{21}=-i(1+i\barSigma_{12})^{-1}(1+\barSigma_{11}\barG_{11})$.
If we introduce this expression for $\barG_{21}$ into
equation \eqref{upperleft}, rearrange a bit and
use the operator identity 
$1+O(1-O)^{-1}=(1-O)^{-1}$, we obtain
\begin{eqnarray*}
(1+i\barSigma_{21})\barG_{11} &=& (\Gzero-\barSigma_{22})
  (1+i\barSigma_{12})^{-1}(1+\barSigma_{11}\barG_{11}).
\end{eqnarray*}
Hall's equation is recovered by identifying 
$A=-i\barSigma_{21}$,
$B=-i\barSigma_{12}$ and $C=-\barSigma_{22}$, where the right
hand side is taken at $k_+=k_-=0$.
Note that Hall's equation is now obtained after a few lines
of algebra instead of a subtle analysis of the graphical
structure of the diagrams.

With the same approach, all the nonperturbative methods
used in solid-state physics, such as the 
GW approximation \cite{GW} and the Bethe-Salpeter
equation \cite{Albrecht,Benedict}, can
be transposed to the case of a general
initial state. This will be presented in
a forthcoming publication.

\section{Determination of the ground state}

QFT with a general state was studied because the 
initial eigenstate of a quantum system is sometimes
degenerate. However, it remains to determine
which density matrix $\omega_{mn}$ of the 
free Hamiltonian leads to the ground state
of the interacting system.

A solution to this problem was inspired by quantum
chemistry methods \cite{BrouderICDIM}.
A number of eigenstates $|\Phi_n\rangle$ of $H_0$
are chosen, for example the complete list of degenerate
eigenstates corresponding to a given energy.
These eigenstates span the so-called \emph{model space} and
the ground state of the interacting system is assumed
to belong to the adiabatic evolution of the model space.
This model space generates, for each density matrix,
a linear form $\omega$ 
as described in equation \eqref{evAomega}. The problem
boils down to the determination of the density matrix
$\omega_{mn}$ that minimizes the energy of the interacting
system.

This minimization leads to an effective Hamiltonian
and the proper density matrix is obtained by
diagonalizing the effective Hamiltonian.
This type of method is typical
of atomic and molecular physics \cite{LindgrenMorrison}.
However, the effective Hamiltonian can now be determined
by powerful non-perturbative Green function methods.
Therefore, the present approach leads to a sort of unification
of quantum chemistry and QFT: it contains
standard QFT when the dimension of the model space is one,
it contains standard quantum chemistry (more precisely
many-body perturbation theory) when the Green functions
are expanded perturbatively.

Therefore, the present approach might help developing
some new nonperturbative methods in quantum chemistry.
On the other hand, quantum chemistry has accumulated
an impressive body of results. The physics Nobel-prize
winner Kenneth Wilson stated that \cite{KWilson}
``Ab initio quantum chemistry is an emerging computational
area that is fifty years ahead of lattice gauge theory.''
Therefore, the experience gained in quantum chemistry
can be used to solve some of the remaining problems of the present approach,
such as the removal of the secular terms\cite{Kukharenko} to all
order.

\section{Conclusion}

The present paper sketched a new method to determine the
Green functions of quantum field theory with a general state.
The main idea is to transform the cumulant function describing
the intial state into an interaction term. 
As a consequence, the cumulants become dressed by
the interaction, providing a much better description of the
correlation in the system.

An alternative method would be to work at the operator level,
as was done recently by D\"utsch and Fredenhagen \cite{Dutsch04},
and to take the expectation value at the end of the calculation.
This would have the obvious advantage of dealing with a
fully rigorous theory. However, we would loose the non-perturbative
aspects of the present approach.

Although this approach seems promissing, much remains to be
done before it can be applied to realistic systems: 
(i) our description is purely formal; (ii) the 
degenerate initial eigenstates lead to secular terms that must
be removed \cite{Kukharenko}; (iii) renormalization
must be included, although this will probably not be
very different from the standard case, because all the 
singularities of the free system are restricted to $\Gzero$.

Interesting connections can be made with other problems.
For example, the cancellation theorem \cite{Tikhodeev}
seems to be interpretable as a consequence of
the unitarity of the S-matrix. It would
extend Veltman's largest time equation \cite{Veltmancut}
to the case of spacetime points with equal time.
Another exciting track would be a connection with noncommutative
geometry. Keldysh\cite{Keldysh} noticed that the doubling of sources could
be replaced by a doubling of spacetime points. In other words,
$j_\pm(x)$ becomes $j(x_\pm)$, where $x_\pm$ are two copies of the
spacetime point $x$: time travels from the past
to the future for $x_+$ and in the other direction for $x_-$.
Sivasubramanian and coll. \cite{Sivasubramanian} have
proposed to interpret this doubling of spacetime points 
in terms of noncommutative geometry. It would be interesting
to follow this track for our quadrupling of spacetime points.

From the practical point of view, the main applications of 
our scheme will be for the
calculation of strongly-correlated systems, in particular
for the optical response of some materials, such as gemstones,
that remain beyond the reach of the standard tools of contemporary
solid-state physics.

After the completion of this work, we came across a little known article
by Sergey Fanchenko, where the cumulants are used to define 
an effective action \cite{Fanchenko}. His paper is also
interesting because it gives a path integral formulation
of quantum field theory with a general state.
His approach and the one of the present paper provide
complementary tools to attack nonperturbative problems 
of quantum field theory with a general state.



\subsection*{Acknowledgment}
I thank Alessandra Frabetti, Fr\'ed\'eric Patras, Sergey Fanchenko and Pierre
Cartier for very useful discussions.

\end{document}